\documentclass{llncs}
\usepackage[latin1]{inputenc}
\usepackage{cite}
\usepackage{subfigure}
\usepackage{listings}
\usepackage{amsmath,amssymb}
\usepackage{multirow}
\usepackage{multitoc}
\usepackage{rotating}
\usepackage{epsfig}
\usepackage{xcolor}
\usepackage{enumitem}

\makeatletter
\let\@copyrightspace\relax
\makeatother

\title{A Conceptual Model for Context Awareness in  Ethical Data Management}

\author{
Elisa Quintarelli\inst{1} \and Fabio Alberto Schreiber\inst{2} \and Kostas Stefanidis\inst{3} \and Letizia Tanca\inst{2} \and Barbara Oliboni\inst{1}
}
\authorrunning{E. Quintarelli et al.} 
\institute{
University of Verona, Italy \\
\email{\{elisa.quintarelli,barbara.oliboni\}@univr.it}
\and
Politecnico di Milano, Italy\\
\email{\{fabio.schreiber, letizia.tanca\}@polimi.it}
\and
Tampere University, Finland\\
\email{konstantinos.stefanidis@tuni.fi}
}

\begin{document}

\maketitle

\thispagestyle{empty}
\begin{abstract}

Ethics has become a major concern to the information management community, as both algorithms and data should satisfy ethical rules that guarantee not to generate dishonourable behaviours when they are used. However, these ethical rules may vary according to the situation-the context-in which the application programs must work. 
In this paper, after reviewing the basic ethical concepts and their possible influence on data management, we propose a bipartite conceptual model composed of the Context Dimensions Tree (CDT), which describes the possible contexts, and the Ethical Requirements Tree (ERT), representing the ethical rules necessary to tailor and preprocess the datasets that should be fed to Data Analysis and Learning Systems in each possible context. We provide some examples and suggestions on how these conceptual tools can be used.
\end{abstract}

\section{Introduction}

History tells us that, after the blow of the atomic bombs in Japan, the physicists \textit{lost their innocence}. Now it comes to Computer Science: the recent developments of Artificial Intelligence (AI), both in theory and applications, have raised a lot of concerns about ethical issues \footnote{Other concerns involve the ecological impact of the Machine Learning algorithms, e.g., the energy needed for processing Neural Networks \cite{AIcost2023}.}. Every discipline, in both technical and social domains, has its behavioral codes, established and monitored by professional organizations,  associations, and civil society. AI is no exception, but its behavioural code has not been fully specified yet: ethics in the context of Machine Learning (ML) and, in general, in the use of Data Science to make decisions, still raises perplexities that are far from being solved. 

An ethical behaviour can be obtained by applying rules that implement appropriate guidelines that should be followed. In the early fifties of the last century, a seminal paper by Norbert Wiener~\cite{wiener_1950, bynum_2008} addressed the ethical issues related to ``cyber-technologies". Several documents, issued by professional and regulatory institutions~\cite{SSESC2021, ACMce2018, USic20xx, Fossa}, formulate deontological rules that should be obeyed, and many papers, \cite{Wasabi2023, S&P2021, Vardi2022, Online21, CACM2023, professional} among them,  argue about what to do and what not to do in the development and use of information systems.  

On the other hand, different boundary conditions may require different behaviours. As an example, a widely recognized rule states that-in most applications-the  {\it gender}, or the {\it age}, should not be used to discriminate among people; however, a well-established behaviour during accidents (such as, e.g., a shipwreck) states {\it ``women and children first"}, which, although motivated by the need to perpetuate the human species, appears as a discrimination against men. In contrast, several recruiting procedures favour men, based on the prejudice that men are more available than women, and, even on this topic, some movements argue that fighting discrimination against women adversely affects men.

Probably, we all agree with the great philosopher Immanuel Kant, who conceived the moral law as a \textit{categorical imperative}~\cite{kant_1785} - i.e. ``an absolute, unconditional requirement that must be obeyed in all circumstances and is justified as an end in itself''. However, the examples above clearly show that ethical rules can depend on the ambient and circumstances in which they must be applied, i.e., they are \textit{context-aware}~\cite{Demartini2024,vehicle21,vehicle22, 5models, decision}.

In ordinary life, human beings know how to recognize the context in which they find themselves and consequently how to guide their actions or conversations. This ability does not transfer automatically to computers; therefore, when decisions are made based on Machine Learning algorithms, or, in general, on Data Science methods, ethical problems may originate from the data provided as input. 
Decisions are made based on previous experience and data, and much of this data comes from -- possibly old -- information systems and databases, possibly reflecting prejudices and preconceptions of that time that clash with current ethical principles. 
Thus, data used to support decisions or fed to machine learning systems should be carefully controlled to obtain ``tailor-made" solutions that are appropriately balanced \emph{with respect to the specific context of use}~\cite{CommunicACM2024, Bjornsson2013-BJRCIE}.

According to Dey~\cite{dey}, \textit{context is any information that can be used to characterize the
situation of an entity. An entity is a person, place, or object that is considered relevant to the interaction between a user and an application, including the user and the application itself.} 

It is necessary to devise a model that can represent the possible contexts and a mechanism to ``teach" computers how they ``should behave" in each context. This would enable us to trigger the appropriate ethical rule when hiring people or when saving people during a shipwreck. 

\smallskip

The main contributions of this paper are the following:
\begin{itemize}
    \item 
    We propose the use of a tree-based \textit{conceptual model}~\cite{IJWET2007,bolchiniCDT} to represent  the various possible situations (e.g., recruitment process) as well as the possible ethical aspects of data in a \emph{context-aware} manner. Indeed, ethics also has various dimensions~\cite{Firmani2020}, such as fairness, diversity, privacy, and transparency, and the conceptual representation we propose    describes both the possible context dimensions and ethical dimensions of a target scenario.
    \item 
    We propose a \textit{methodology} to reveal, in the dataset to be used for  analysis or for system training, the data that are objectionable \emph{w.r.t. the ethical behaviour that should  be applied in a certain context}, helping the designers to transform those data in order to reduce possible underlying  discriminations.  
\end{itemize}

The paper is organized as follows: Section \ref{ethicaldimension} introduces various ethical dimensions, while Section \ref{datatrasformation} deals with  state-of-the-art  methods for modifying the data and obtaining an ethical behaviour in ML or  decision systems. 
Section \ref{model} presents a tree-based conceptual model to jointly describe contextual information and ethical dimensions, and Section \ref{trasformation} outlines our proposal for context-aware ethical data transformations. 
Finally, Section 6 concludes the paper with a summary and directions for future research.

\section{Ethical Dimensions}
\label{ethicaldimension}

Ethical requirements can become dominant and change with time or in different cultures.  
The main dimensions of Data Ethics~\cite{Firmani2020} can be described as follows:

\textbf{Fairness:} Fairness of the data is defined as \emph{lack of bias}~\cite{stoyanovich2016data}. Its importance has been acknowledged, due, for instance, to the unsettling consequences of training systems with biased data~\cite{floridi2018ai4people}. 
Fairness does not require discrimination against certain elements based on attributes values that are not relevant to the task at hand, such as \emph{gender, religion, age, sexual orientation} and \emph{race}, attributes called \emph{protected}, or \emph{sensitive}. 

At a high level, we can distinguish between two approaches to formalize fairness: (i) \textit{individual fairness}, based on the premise that similar entities should be treated similarly, and (ii) \textit{group fairness}, which asks that all groups be treated similarly.
 
Human bias generally represents the tendency to prefer someone/something to another person/thing. Algorithmic bias refers to the tendency of algorithms to reflect human biases by systematically discriminating against some (groups of) individuals in favor of others. The biased algorithm outcomes arise from erroneous assumptions of the process, which are strictly related to the data used as input or to train the algorithm.
  
Another aspect of fairness concerns the concepts of \textit{Equality} and \textit{Equity}, which, in the colloquial language, are used as synonyms, but are not exactly the same. In fact, we speak about Equality when each person or group is given \textit{equal resources or opportunities}, while Equity recognizes the specific circumstances of each person or group, and treats people differently depending on their endowments and needs (focuses on equality of outcome)~\cite{osti_10287321}. Note that there are many different and interesting definitions for all these concepts, and, for the moment, we will not address all the subtleties and distinctions.

\textbf{Transparency:} When considering ethical issues, transparency means working in such a way that others can see which actions are performed. Transparency provides a complete report of information, facts, and contextual situations to guarantee an ethical decision-making process. Therefore, transparency is the ability to interpret the information extraction process and verify which aspects of the data and procedures determine its results. 

In this context, transparency metrics can be based on concepts such as (i)
\emph{data provenance}~\cite{GlavicD07,stoyanovich2016data,simmhan2005survey}, which includes information about the processes and source data items that lead to its creation and current representation, and (ii) \emph{explana\-tions}~\cite{rader2018explanations}, which describe how a result has been obtained. In particular, explanations refer to the ``interpretability'' of results produced by algorithms, procedures, and machine learning techniques. This means that it must be possible to explain \emph{why a given model is used and why a given output is returned}.

\textbf{Diversity:} Diversity is the degree to which different kinds of objects are represented in a dataset. Several metrics are proposed in~\cite{drosou2017diversity}. Ensuring diversity at the beginning of the information extraction process may be useful to enforce fairness at the end. The diversity dimension may conflict with established dimensions in the \emph{Trust} cluster of~\cite{DBLP:series/dcsa/BatiniS16}, which prioritizes a few high-reputation sources overall.

\textbf{Data Protection:}  The data-related aspect of \emph{Privacy} refers to the proper handling of sensitive information, such as personal data and confidential data. Personal data (i.e., Personally Identifiable Information) is information that can be used to identify an individual (e.g., name, social security number, place, and date of birth). Confidential data represents secret and private information and includes medical records, a person's address and phone number, and financial records. Privacy problems may arise from the fact that different anonymized datasets might reveal sensitive information when combined~\cite{tockar2014riding}. 

Note that sometimes conflicts can arise among the various ethical dimensions. As an example, privacy requirements might conflict with those for transparency and explainability, like in the case of medical needs or when selecting the appropriate person for an official position in a country where religious conflicts are raging.

\section{Context-oblivious Ethical Data Transformation: \\Basic State of the Art} 
\label{datatrasformation}
In this section, we survey some classical, context-oblivious methods for transforming the data that has to be used as input to a decision system to obtain an ethical behaviour, e.g., by removing any underlying bias or discrimination. Typically, such methods are considered application-ag\-no\-stic, and their goal is to mitigate the effects of the presence of non-ethical behaviours in the training data ~\cite{catania2022}. 

As already noticed, the bias in input datasets may be due to old discriminatory habits or to the process used for collecting them, like, for example, the assumptions  made for the missing-value imputation or for the pieces of training data to be collected. 
In the following, we consider some situations and show how taking the context into account can support the choice of the appropriate ethical data transformation.

\textbf{Data Suppression:} 
 
A typical baseline solution~\cite{DBLP:journals/kais/KamiranC11}   to tackle bias in the data is to simply \textit{omit specific attributes} when training a classifier: the idea here is to identify the attributes that correlate most with the sensitive attributes, then remove both the sensitive attributes and the most correlated ones to reduce the effect of the interdependence between the class labels and the sensitive attributes.

\textbf{Database Repair:} 

Rather than by eliminating whole columns, non-ethical behaviours in data can be avoided by modifying (or even removing) information from the sensitive attributes. 
Towards this direction, \cite{DBLP:conf/sigmod/SalimiRHS19} presents an approach that removes discrimination by \textit{repairing the training data to remove the effects of discriminatory causal relationships between the sensitive attribute and the  predictions of the classifier}. 

Instead, \cite{DBLP:conf/kdd/FeldmanFMSV15} uses a test for \emph{disparate impact}, in accordance with how well the sensitive class can be predicted based on the other attributes: a repair solution for numerical attributes is applied, which allows changing the data so that prediction of the class is still possible. 

Moreover, \cite{DBLP:journals/corr/LumJ16} uses a chain of conditional models to protect and adjust attributes of arbitrary number and type (e.g., attributes that may be continuous or discrete), so that statistical models for handling data sparsity, such as Bayesian hierarchical models, can be deployed.

\textbf{Data Transformation:} 

The goal of \cite{DBLP:conf/nips/CalmonWVRV17} is to balance discrimination with the utility of processed data, formulating this as an optimization problem to produce transformations with three goals: \textit{controlling discrimination, limiting distortion in individual data samples,} and \textit{preserving utility}. 

Discrimination control aims to limit the dependence of the transformed outcome on the protected variables, which requires the conditional distribution of the outcome to the sensitive variables to be close to a target distribution for all sensitive values, while distortion control should satisfy constraints that avoid large changes. 

Finally, utility preservation should ensure that a model learned from the transformed data (when averaged over the protected variables) is not too different from one learned from the original data. 

\textbf{Data Augmentation:} 
 
Rather than transforming existing data under the umbrella of recommender systems, \cite{DBLP:conf/wsdm/RastegarpanahGC19} proposes to augment the input with additional data to improve the resulting recommendations. Motivated by the fact that \textit{adding new input data may be easier than modifying existing data inputs}, especially when a system is already running, this framework starts from an existing matrix factorization recommender that has already been trained with some input data and considers adding new users who provide ratings for existing items. The new ratings are chosen in such a way as to improve the socio-technical aspects of the recommendations.

\textbf{Massaging the Dataset:} 

To remove discrimination from input data, the approach of ~\cite{DBLP:journals/kais/KamiranC11} changes the labels of some data objects in the dataset. Since it is crucial to identify which labels should be changed, \textit{a ranking method is applied to select the best candidates for relabeling}. Extending some ideas proposed in~\cite{DBLP:conf/icdm/CaldersKP09}, massaging considers a subset of data from protected and unprotected groups as promotion and demotion candidates, respectively, and changes their class labels. The data objects are ranked based on their probability of having positive labels, and the top-k protected objects are chosen for promotion, while the bottom-k unprotected ones are chosen for demotion.

\textbf{Reweighting:} 
 
With this approach, rather than changing the object labels, a weight is assigned to the tuples~\cite{DBLP:journals/kais/KamiranC11}. This will allow us to \textit{give more importance to some attribute values than to others}. In case the possible values of the important attribute are more than 2, and we want to favor the tuples presenting more than one of those values, the tuple weight must be the product of the weights assigned to each such value.

\section{The Context \&  Ethical Dimension Models}
\label{model}

The notion of context in computer science was introduced by Dey in 2001~\cite{dey}, and a lot of work has been done since then~\cite{data:oriented:survey}. 

In this paper, we will introduce a model and a methodology to tailor the data used as input to a decision system to obtain, for each possible context of use of a certain dataset, the most appropriate ethical behaviour.

As a conceptual model for representing contexts, we adopt the Context Dimension Tree (CDT), formally defined in~\cite{IJWET2007,bolchiniCDT}. 
To explain the CDT model, we refer to an example:  consider the work domain, where employees may belong to public or private institutions and have a role within some specific job, possibly entailing some risks. The persons' financial and family status are also recorded, along with some other data.
Fig.~\ref{tab:rel} shows a fragment of a relational database useful in the described scenario.

\begin{figure}[!htb]
\begin{center}
\begin{tabular}{|l|}
\hline
\\
\texttt{
\small{PERSON(\underline{pID}, Surname, Name, DateOfBirth, Gender, Citizenship,}}\\ 
\texttt{
\small{\hspace{0.9cm} FamSituation, Census, Ethnicity, JobType, \dots)}}\\
\texttt{
\small{PREGNANCY(\underline{WomanId}, \underline{IniDate}, FinDate)}}\\
\texttt{
\small{INSTITUTION(\underline{InstName}, Type, Domain, Location, CurrentSituation, \dots}}\\
\texttt{
\small{EMPLOYEE(\underline{InstName}, \underline{pID}, \underline{Role}, IniDate, Department, Performance)}}\\
\texttt{
\small{OFFERS(\underline{InstName}, \underline{JobType})}}\\
\texttt{
\small{RISKS(\underline{RiskName}, \underline{JobType})}}\\
\texttt{
\small{INCURS(\underline{pID}, \underline{RiskName}, GravityLevel)}}\\
\texttt{
\small{ROLE-SALARYS(\underline{Role}, \underline{Seniority}, \underline{JobType}, Salary)}}\\
\texttt{
\dots}\\
\\
\hline
\end{tabular}
\end{center}
\caption{A simple Relational Database schema}
\label{tab:rel}
\end{figure}

\begin{figure}[htpb]
   \includegraphics[width=\linewidth]{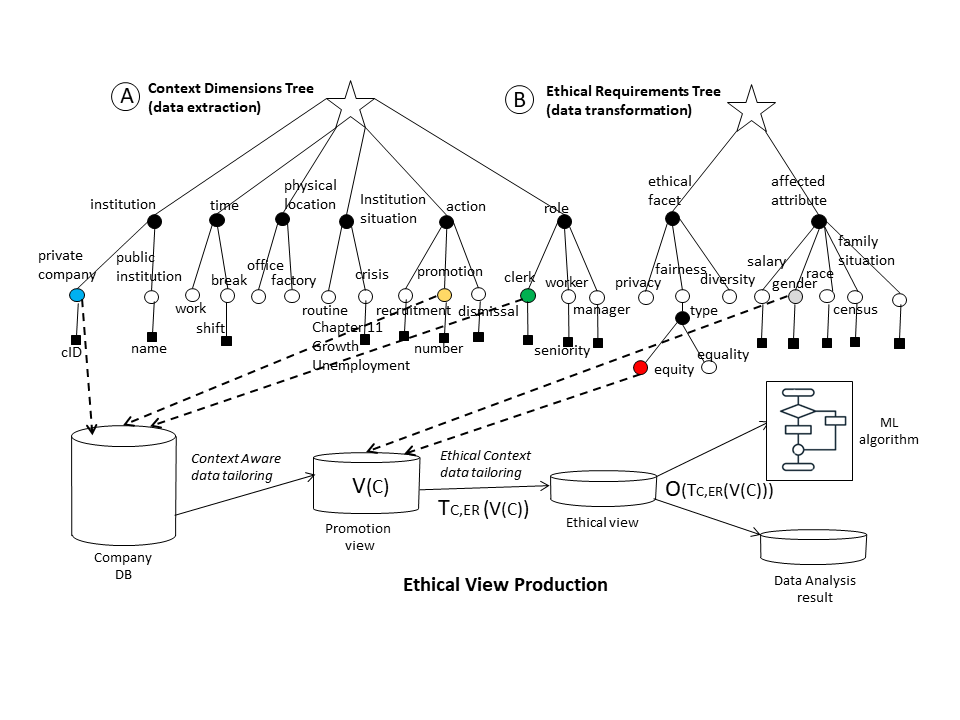}
    \vspace{-1.8cm}
    \caption{CDT (A), ERT (B) and Ethical View Production for a personnel management application }
    \label{figure.personnel}
\end{figure}

In synthesis, a CDT (see part (A) of Fig.~\ref{figure.personnel}) is built as follows:
\begin{itemize}[leftmargin=*] 
 
\item   A \textbf{star-shaped, white} node is the \textit{\textbf{CDT root}}. 
 
\item   The children of the root are \textbf{black} nodes, which represent the context \textit{\textbf{dimensions}}, i.e., the perspectives relevant to select, from the database of Fig.~\ref{tab:rel}, the data that are relevant in each context.
 
\item Each dimension node can have as children one or more \textbf{white} nodes, called \textit{\textbf{concept nodes}}, which represent the \textbf{values} the dimension can assume.  As an example, in Fig.~\ref{figure.personnel}(A), a dimension is the \textbf{role} of a person, with three white nodes ``worker", ``clerk," and ``manager" as possible values. 
 
\item Concept nodes can feature one or more \textit{\textbf{subdimensions}}, also represented as  \textbf{black} nodes, and the same structure can be repeated up to the required level of detail. 

 \item  Both dimension and concept nodes can have \textit{\textbf{attributes}}, represented as small \textbf{square} nodes. An attribute attached to a dimension node \textit{is a shorthand adopted when the possible values of that dimension are many}\footnote{E.g., if the number of possible values of the dimension  \textbf{role} had been high, we could have represented them by replacing the three white nodes with an attribute.}. 
 Instead, an attribute attached to a white node \textit{represents the possible values that the white node can assume}\footnote{E.g., in Fig.~\ref{figure.personnel}, the name of the Public Institution is represented by an attribute}.    
\end{itemize}

\vspace{-0.9cm}
\smallskip

A {\it context} $C$ is represented by {\it a tuple of context elements}, each  described in the form
\emph{dimension name = value}, where, if the white node has an attribute, its value is also included in the \emph{value} part. 

In a context, sibling white nodes are mutually exclusive since they represent orthogonal concepts, while sibling black nodes represent the different (sub)di\-men\-sions that define a concept.

The main dimensions of Fig.~\ref{figure.personnel}(A) represent the different context perspectives: \textit{the black node {\bf action} specifies the purpose of the data analysis}, while the other black nodes are the dimensions that describe the people and the organizations involved in the choices that should be made, along with their peculiarities (if any).

We adopt a similar model to represent the ethical facets, i.e., the perspectives that drive the ethical transformation. Specifically, Fig.~\ref{figure.personnel}(B) shows an Ethical Requirements Tree (ERT), whose two black nodes represent:
\begin{itemize}[leftmargin=*]

\item The {\bf ethical facets} related to the situation at hand: in this example, we consider {\it privacy, fairness}, and {\it diversity}. Note also the further specialization of the {\it fairness} facet into the two values {\it equity} and {\it equality}. 

\item The {\bf affected attribute} is the attribute of the input dataset $D$ for which we want to guarantee compliance with the chosen ethical dimension(s). \textit{The values (i.e., white nodes) of this dimension are the sensitive attributes that can help determine the most suitable ethical behavior in a certain CDT context.} 
\end{itemize}

The lowest level of Fig.~\ref{figure.personnel} shows the parts of the initial Database that are affected by the transformations guided by the two trees. From the Company Database, on the left, we select the data, called Contextual View, that is to be used according to the reference context (the Promotion View in the context highlighted by Fig.~\ref{figure.personnel} (A)). 
Analyzing the Contextual View may reveal the need to adopt ethical behaviors, possibly with different priorities, to prevent unfair outcomes when applying Analysis/ML algorithms. Indeed, moving to the right, the Ethical View is obtained on the basis of the procedure envisaged by the selected ethical facet highlighted in Fig.~\ref{figure.personnel}(B), which is further explained in the next section. At this point, the Ethical View will be used to feed the Analysis/ML algorithm.
The next Section will illustrate all these steps with an example that shows how this conceptual model should be used in various circumstances.

\section{Context-aware, Ethical Data  Transformation}
\label{trasformation}

Our methodology proposes a structured approach for context $C$, beginning with an analysis of the Contextual View ${V}(C)$ defined as a set of relational algebra expressions highlighting the data relevant for context $C$ and that may reveal potential imbalances. 
Based on this analysis, we determine the most appropriate ethical value $ER$ of the ethical facet, useful for the transformations $T_{C,ER}$ to be applied to the view ${V}(C)$,  before the specific data science operation $O$, (e.g., a query, statistical operations, an ML algorithm). The goal is to ensure that $O(T_{C,ER}({V}(C)))$ satisfies both the requirements for the context $C$ and the ethical criteria forced by $ER$. 

Note that in our setting, the operation $O$ might be just a top-k query that could retrieve the set of the new managers or, instead, the training of a machine-learning (ML) model that will be learned from this data and adopted in the future to choose the applicants that should become managers.

As explained, we consider as an example the procedure for promoting to the role of manager some clerks of a private company, whose data are stored in the Company DB.
 
The first step consists of creating the Contextual View - the Promotion view -  \textit{that contains all and only the data relevant to} the promotion procedure in the company (e.g., number of clerks to be promoted, data on the managers already in service, clerk performance and personal data, etc.)[4]. 

Fig.~\ref{figure.personnel} (Ethical View Production) shows the flow when the context related to the action ``promotion" (yellow-colored node of the CDT) of clerks (green-colored node of the CDT) to managers combined with the ``equity" Ethical Requirement (red-colored node) w.r.t. the ``gender" attribute (grey-colored node). 

From now on, let us call \textit{Ethical Context} the union of the attributes chosen from the CDT and the ERT.

Formally, the contextual view related to the context $C$ in a target company $cID$ can be described as: 
\begin{align*}
V(C)= \{ &
E_1\equiv\mathtt{\sigma_{Role=``clerk"}(EMPLOYEE \bowtie PERSON)}, \\
& E_2\equiv\mathtt{\sigma_{Role=``manager"}(EMPLOYEE \bowtie PERSON)} \}
\end{align*}

Ideally, promotions should be based on the performance data of clerks; thus, when considering {\it equality}, the Ethical View can be obtained by ordering $E_1\in V(C)$ based on the $\mathtt{Performance}$ attribute.   

However, if the principle of \textit{equity} is to be applied, additional factors and corresponding actions must also be considered.
A preliminary analysis of the Promotion View guided by the possible affected attribute values (i.e., gender, race, salary, \dots) may suggest the need for a rebalancing w.r.t. one (or more) attribute(s). We remark that, without loss of generality, the affected attributes can be considered categorical attributes.
For all affected attributes $A_i$ with domain $Dom(A_i)$, we group the tuples in $V(C)$ to compare the cardinalities of the resulting groups. In the extended relational algebra with $\gamma$ for expressing the grouping criteria, we first specify the groups as follows: 
\begin{align*}
\{ & E_1'\equiv\gamma_{A_i}\mathtt{\sigma_{Role=``clerk"}(EMPLOYEE \bowtie PERSON)}, \\ & E_2'\equiv\gamma_{A_i}\mathtt{\sigma_{Role=``manager"}(EMPLOYEE \bowtie PERSON)}\}
\end{align*}

Given an affected attribute $A_i$, if there are at least two of its groups with greatly different cardinalities, the analysis suggests a disparity between the groups. For instance, when considering $A_i=gender$, if a disparity between male and female managers emerges when grouping $E_2$, a gender rebalancing is suggested by the promotion context $C$, which requires a \textbf{DATABASE REPAIR} action. 

A possible solution is \textit{ to supplement the training dataset with an appropriate number of duplicates of female employees}. 
Given the Best Male Clerks $BMC$  
and the Best Female Clerks $BFC$ by selecting only the tuples of $E_1$ with Performance $>Pmin$, formalized in relational algebra as follows
$$BMC\equiv\mathtt{\sigma_{Role=``clerk" \wedge Gender=``male"\wedge Performance >Pmin} (EMPLOYEE \bowtie PERSON)}$$
$$BFC\equiv\mathtt{\sigma_{Role=``clerk" \wedge Gender=``female"\wedge Performance >Pmin} (EMPLOYEE \bowtie PERSON)}$$

we should add 
$\lceil(BMC-BFC)/BFC\rceil$ copies of $BFC$ to $BMC \cup BFC$, in order to have the same percentage of genders, and submit the result to the final ML algorithm.

\medskip
\noindent
Thus, to obtain the Ethical view $T_{C,ER}({V}(C))$, we have to replicate $p=\lceil (BMC-BFC)/BFC\rceil$ times the tuples of BFC:

\medskip
\noindent
{\it repeat $p$ times} $\{$

\noindent
\textbf{INSERT INTO } $EV(C)$ \textbf{ AS } \\
\textbf{   SELECT } * \\
\textbf{   FROM } BFC  $\}$

\medskip
\noindent
In this way, the table $EV(C)$ is balanced w.r.t. the gender attribute and can be used as the input $O(T_{C,ER}({V}(C)))$ for an ML algorithm that suggests the clerks to be promoted. 

This method balances the candidates on the basis of the current percentages of Best Clerks; however, the balance might also be based on the current percentages of Managers.
Let $MM$ be the number of Male Managers and $FM$ the number of Female ones; then, we should add 
$\lceil(MM-FM)/(MM+FM)\rceil$ copies of $BFC$ to $BMC \cup BFC$ to the dataset fed to the ML algorithm.

In a slightly different context, for example in the context $C_4$ below, related to the recruitment of managers when the affected attribute is \textbf{race}, we could obtain a contextual view where the race distribution is suppressed for the Privacy Ethical Requirement.

In the following, we list some more Ethical Contexts, and the related Ethical Transformations, for an ethical personnel management application. Note that the contexts $C$ and $C_{1b}$ refer to the same Action but, since the chosen type of Fairness is different, they need different repair Actions.

\begin{itemize}[leftmargin=*]
\small
\item \textbf{DATA SUPPRESSION}.\\
    $C_{1b}:$ \textit{ACTION = promotion} to manager, \textit{ROLE = clerk, \\ETHICAL DIMENSION = \textcolor{red}{equality}, \\ AFFECTED ATTRIBUTE = gender}.\\
    Delete the gender column and all the other columns that may suggest the gender value (e.g., number of pregnancies). 
\medskip    
\item \textbf{REWEIGHTHING}.\\
    $C_{2}:$ \textit{ACTION = promotion, ROLE = worker, \\ETHICAL DIMENSION = \textcolor{red}{equity}, 
    \\ AFFECTED ATTRIBUTE = family situation}.\\
   Other conditions being equal, priority should be given to applicants with family difficulties (e.g., many children, widowed). Assign a weight to each difficulty and replicate the various groups corresponding to each difficulty by multiplying them by the weight. 
\medskip 
\item \textbf{DATABASE REPAIR}.\\
    $C_{3}:$ \textit{ACTION = dismissal, ROLE = clerk, \\ETHICAL DIMENSION = \textcolor{red}{diversity}, \\ AFFECTED ATTRIBUTE = gender}.\\
    Other conditions being equal, dismiss more people of the numerically dominant gender.    
\medskip 
\item \textbf{DATA SUPPRESSION}.\\
    $C_{4}:$ \textit{ACTION = recruitment, ROLE = manager, \\ETHICAL DIMENSION = \textcolor{red}{privacy},\\ AFFECTED ATTRIBUTE = race}.\\
    Delete the race column and all the other columns that may suggest the race value. 
\end{itemize}

\paragraph{\textbf{Discussion}.}
\label{discussion}

Although it is clear that equity and equality pursue different goals and thus are selected as alternative ethical dimensions, balancing other ethical facets, or giving them priority, requires careful design in contexts when multiple behaviours could be adopted.

For the above-mentioned examples, when dealing with equity, sensitive information (e.g., gender, family situation), usually protected by the privacy principle, is essential for responsible practices; thus, equity has a priority on privacy, but to preserve, when possible, other sensitive information, only the minimal amount of sensitive information should be revealed.

When underrepresented groups are present in a reality (e.g., a minority of women as managers during a promotion process) the equity principle is suggested. However, ensuring diversity by adding a new role at the management level could add value, but only when it does not come at the expense of fairness.

In our example about promoting some clerks to the role of manager, we might want to consider both fairness and diversity, with different priorities.
Each principle will impact a different attribute, such as \emph{gender} for equity and \emph{department} for diversity, and thus produce a different transformation of the Promotion View.
In this case, we can fix a priority in evaluating the different facets by defining percentages, e.g.: $pr_{eq}=60\%$ and $pr_{div}=40\%$.

A possible approach for taking into account both aspects would require evaluating each principle in order of priority (in this case, first equity and then diversity), on a corresponding percentage of available positions before rebalancing the dataset.
Let $N$ be the number of promotions planned in the company. We will apply the equity principle on $60\%$ of $N$, and the diversity principle on the remaining $40\%$. It should be noted that for small values of $N$ involving, as in this example, non-divisible positions, it may not be possible to split the set of available positions. In this case, the result shall be rounded to the nearest reasonable integer.

\section{Conclusions and Future Work}

The focus of this work is on defining a conceptual model to represent both the contextual and ethical aspects of data. To describe the notion of context, we adopt the Context Dimension Tree, which represents the various contexts by using a set of dimensions that take different values according to each specific situation. We follow a similar way, that is, a tree structure, to describe ethics and its dimensions, such as fairness, diversity, privacy, and transparency. 

Our proposal takes into account existing context-oblivious methods for tailoring the data that have to be used as input to a decision system to obtain an ethical behaviour (for example, by removing any underlying bias or discrimination), and 
suggests a methodology to take into consideration the contextual requirements when deciding which specific method should be applied. 

In this paper, we have just begun to realize the need for ethical and context-aware data transformations. Next, we highlight a critical open issue and challenge for future work. A problem intrinsic to all definitions of ethics in data management, especially when considering different contexts, is the fact that they attempt to quantify philosophical, legal, often elusive, and even controversial notions of justice and social good. Unfortunately, when notions that reflect value systems and beliefs interact with the mechanisms to measure and implement them, the complexity increases. 

Clearly, there should be a way to distinguish between what constitutes a belief and what is the mechanism, or measure, for codifying this belief: from a technical point of view, we should be able to focus on assessing whether a proposed measure is an appropriate codification of a given belief instead of assessing the belief itself. Future work calls for the development of different levels of abstraction and mappings between them.

\section*{Acknowledgements} 
We wish to thank the students L. Liparulo, F. A. Mazzola and  N. Tummolo for the useful discussions and experiments.

\bibliographystyle{abbrv}
\bibliography{bibetica,mainJDIQ}

@inbook{bynum_2008, place={Cambridge}, series={Cambridge Studies in Philosophy and Public Policy}, title={Norbert Wiener and the Rise of Information Ethics}, DOI={10.1017/CBO9780511498725.002}, booktitle={Information Technology and Moral Philosophy}, publisher={Cambridge University Press}, author={Bynum, Terrell Ward}, editor={van den Hoven, Jeroen and Weckert, JohnEditors}, year={2008}, pages={8-25}, collection={Cambridge Studies in Philosophy and Public Policy}}

@inbook{wiener_1950,  title={The Human use of human beings},  booktitle={Cybernetics and society}, publisher={Houghton Mifflin,  (2nd ed. rev.)}, author={Wiener, Norbert}, year={1950-1954}}

@inbook{kant_1785,  title={Grundlegung zur Metaphysik der Sitten, (Groundwork of the Metaphysics of Morals)},  publisher={Cambridge Texts in the History of Philosophy -Cambridge University Press}, author={Kant, Immanuel}, year={1785/1997}}

@inbook{Fossa,  
title={Automi e persone: introduzione all'etica dell'intelligenza artificiale e della robotica},  
booktitle={}, publisher={Carocci}, author={Fossa, Fabio and  Schiaffonati, Viola and Tamburrini, Guglielmo}, year={2021}}

@article{osti_10287321,
 title = {The Many Facets of Data Equity}, url = {https://par.nsf.gov/biblio/10287321}, journal = {theWorkshop Proceedings of the EDBT/ICDT 2021 Joint Conference}, author = {Jagadish, H.V. and Stoyanovich, Julia and Howe, Bill}, editor = {Costa, Constantinos and Pitoura, Evaggelia} }

@inproceedings{GlavicD07,
  author       = {Boris Glavic and
                  Klaus R. Dittrich},
  title        = {Data Provenance: {A} Categorization of Existing Approaches},
  booktitle    = {DBIS},
  year         = {2007}
}

@article{SSESC2021,
  title={IEEE Standard Model Process for Addressing Ethical Concerns during System Design},
  author={Systems and Software Engineering Standards" Committee},
  journal={IEEE Std 7000 - IEEE CS},
  year={2021},
  pages={1-81}
}

@article{Demartini2024,
title= {OPINION: Data Bias Management},
author={Gianluca Demartini and Kevin Roitero  and Stefano Mizzaro},journal      = {Commun. ACM}, year={2024}}

@article{Firmani2020,
 title= {Ethical Dimensions for Data Quality},
  author       = {Donatella Firmani and Letizia Tanca and Riccardo Torlone},
 journal      = {{ACM} J. Data Inf. Qual.},
  volume       = {12},
  number       = {1},
  pages        = {2:1--2:5},
  year         = {2020},
  url          = {https://doi.org/10.1145/3362121},
  doi          = {10.1145/3362121}}

@article{catania2022,
  title={A Coverage-based Approach to Nondiscrimination-aware Data Transformation},
  author={Chiara Acinelli and Barbara Catania and Giovanna Guerrini and Simone Minisi},
  journal={ACM Journal of Data and Information Quality},
  volume={14},
  number={4},
  year={2022},
  pages={1-26}
}

@article{Wasabi2023,
  title={Wasabi: A Conceptual Model for Trustworthy Artificial Intelligence},
  author={Amika M.Singh, Munindar P. Singh},
  journal={IEEE Computer, Feb. 2023},
  year={2023},
  pages={20-28}
}

@article{S&P2021,
  title={Equity and Privacy: More Than Just a Tradeoff},
  author={David Pujol, Ashwin Machanavajjhala},
  journal={Security and Privacy, vol. 19},
  year={2021},
  pages={93-97}
}

@article{professional,
  title={Leveraging Professional Ethics for Resposible AI},
  author={Nicholas Diakopulos and Al.},
  journal={Communications of ACM, vol. 67},
  year={2024},
  pages={19-21},
  doi={10.1145/3625252}
}

@article{Vardi2022,
  title={ACM, Ethics, and Corporate Behavior},
  author={Moshe Y. Vardi},
  journal={Commun. ACM, Vol. 65 No. 3},
  year={2022},
  pages={5}
}

@article{ACMce2018,
  title={ACM Code of Ethics and Professional Conduct},
  author={ACM Code 2018 Task Force},
  journal={https://www.acm.org/code-of-ethics},
  year={2018},
  pages={5}
}

@article{5models,
  title={5 Models for Ethical Decision Making},
  author={People Centric Consulting Group},
  journal={https://peoplecentric.com/blog/leadership/5-models-for-ethical-decision-making/}
}

@article{decision,
  title={Ethical Decision-Making},
  author={People Centric Consulting Group},
  journal={https://serc.carleton.edu/geoethics/Decision-Making},
  year={2024}
}

@incollection{Bjornsson2013-BJRCIE,
	author = {Gunnar Bj\"{o}rnsson},
	booktitle = {The International Encyclopedia of Ethics},
	editor = {Hugh LaFollette},
	publisher = {Blackwell},
	title = {Contextualism in Ethics},
	year = {2013}
}

@article{USic20xx,
  title={Principles of Artificial Intelligence Ethics for the Intelligence Community},
  author={United States Intelligence Community},
  journal={Principles\  of\  AI\  Ethics\  for\  the\  Intelligence\  Community},
  pages={1}
}

@inproceedings{vehicle21,
	Author={Federico Faruffini, Alessandro Correa Victorino, Marie-Helene Abel},
	booktitle={ICHMS},
	Title={Vehicle Autonomous Navigation with Context Awareness},
	Year={2021}
}

@article{vehicle22,
  title={Context-aware behaviour prediction for autonomous driving: a deep learning approach},
  author={Syama R., Mala C. },
  journal={International Journal of Pervasive Computing and Communications},
  year={2022},
  pages={1-14}
}

@article{AIcost2023,
  title={The hidden costs of AI: Impending energy and resource strain},
  author={Deep Jariwala, Benjamin C. Lee},
  journal={https://penntoday.upenn.edu/news/},
  year={2023},
  }

@article{Online21,
 author={Noriega, Pablo and Verhagen, Harko and Padget, Julian and d'Inverno, Mark},
 journal={IEEE Internet Computing}, 
title={Ethical Online AI Systems Through Conscientious Design}, 
year={2021},
volume={25},
number={6},
pages={58-64},
doi={10.1109/MIC.2021.3098324}}

@article{CACM2023,
  title={Ethics as a Participatory and Iterative Process},
  author={Marc Steen},
  journal={Comm. ACM,  Vol. 66 No. 5},
  year={2023} 
}

@article{dey,
 author = {Anind K. Dey},
 title = {Understanding and Using Context},
 journal = {Personal Ubiquitous Comput.},
 volume = {5},
 number = {1},
 year = {2001},
 issn = {1617-4909},
 pages = {4--7},
 doi = {http://dx.doi.org/10.1007/s007790170019},
 publisher = {Springer-Verlag},
 address = {London, UK},
 }

@article{CommunicACM2024,
  title={Inherent Limitations of AI Fairness},
  author={Marteen Buyl and Tijl De Bie},
  journal={Communications of ACM},
  year={2024},
  volume={67},
  number={2},
  pages={48-58}
}

@ARTICLE{IJWET2007,
  author = {Cristiana Bolchini and Carlo Curino and Elisa Quintarelli and Fabio. A. Schreiber and Letizia Tanca},
  title = {Context Information for Knowledge Reshaping},
  journal = {Int. Journal on Web Engineering and Technology},
  publisher = {Inderscience},
  year = {2007}
}

@article{data:oriented:survey,
 author = {Bolchini, Cristiana and Curino, Carlo A. and Quintarelli, Elisa and Schreiber, Fabio A. and Tanca, Letizia},
 title = {A data-oriented survey of context models},
 journal = {SIGMOD Rec.},
 volume = {36},
 issue = {4},
 year = {2007},
 issn = {0163-5808},
 pages = {19--26},
 numpages = {8},
 url = {http://doi.acm.org/10.1145/1361348.1361353},
 doi = {http://doi.acm.org/10.1145/1361348.1361353},
 acmid = {1361353},
 publisher = {ACM},
 address = {New York, NY, USA},
}

@article{bolchiniCDT,
 author    = {Cristiana Bolchini and
               Carlo Curino and
               Giorgio Orsi and
               Elisa Quintarelli and
               Rosalba Rossato and
               Fabio A. Schreiber and
               Letizia Tanca},
 title = {And what can context do for data?},
 journal = {Comm. ACM},
 volume = {52},
 number = {11},
 year = {2009},
 issn = {0001-0782},
 pages = {136--140},
 doi = {http://doi.acm.org/10.1145/1592761.1592793},
 publisher = {ACM},
 address = {New York, NY, USA},
 }

@inproceedings{DBLP:conf/sigmod/SalimiRHS19,
  author       = {Babak Salimi and
                  Luke Rodriguez and
                  Bill Howe and
                  Dan Suciu},
  title        = {Interventional Fairness: Causal Database Repair for Algorithmic Fairness},
  booktitle    = {SIGMOD},
  year         = {2019}
}

@inproceedings{DBLP:conf/kdd/FeldmanFMSV15,
  author       = {Michael Feldman and
                  Sorelle A. Friedler and
                  John Moeller and
                  Carlos Scheidegger and
                  Suresh Venkatasubramanian},
  title        = {Certifying and Removing Disparate Impact},
  booktitle    = {SIGKDD},
  year         = {2015}
}

@article{DBLP:journals/corr/LumJ16,
  author       = {Kristian Lum and
                  James E. Johndrow},
  title        = {A statistical framework for fair predictive algorithms},
  journal      = {CoRR},
  volume       = {abs/1610.08077},
  year         = {2016}
}

@article{DBLP:journals/kais/KamiranC11,
  author       = {Faisal Kamiran and
                  Toon Calders},
  title        = {Data preprocessing techniques for classification without discrimination},
  journal      = {Knowl. Inf. Syst.},
  volume       = {33},
  number       = {1},
  pages        = {1--33},
  year         = {2011}
}

@inproceedings{DBLP:conf/wsdm/RastegarpanahGC19,
  author       = {Bashir Rastegarpanah and
                  Krishna P. Gummadi and
                  Mark Crovella},
  title        = {Fighting Fire with Fire: Using Antidote Data to Improve Polarization
                  and Fairness of Recommender Systems},
  booktitle    = {WSDM},
  year         = {2019}
}

@inproceedings{DBLP:conf/nips/CalmonWVRV17,
  author       = {Fl{\'{a}}vio P. Calmon and
                  Dennis Wei and
                  Bhanukiran Vinzamuri and
                  Karthikeyan Natesan Ramamurthy and
                  Kush R. Varshney},
  title        = {Optimized Pre-Processing for Discrimination Prevention},
  booktitle    = {Advances in Neural Information Processing Systems},
  year         = {2017}
}

@inproceedings{DBLP:conf/icdm/CaldersKP09,
  author       = {Toon Calders and
                  Faisal Kamiran and
                  Mykola Pechenizkiy},
  title        = {Building Classifiers with Independency Constraints},
  booktitle    = {{ICDM} Workshops},
  year         = {2009}
}

@article{simmhan2005survey,
  title={A survey of data provenance in e-science},
  author={Simmhan, Yogesh L and Plale, Beth and Gannon, Dennis},
  journal={ACM Sigmod Rec.},
  volume={34},
  number={3},
  pages={31--36},
  year={2005},
  publisher={ACM}
}

@misc{tockar2014riding,
  url={research.neustar.biz/2014/09/15/riding-with-the-stars-passenger-privacy-in-the-nyc-taxicab-dataset},
  note = {Online; accessed 25 Nov. 2019}
 }

@inproceedings{rader2018explanations,
  title={Explanations as mechanisms for supporting algorithmic transparency},
  author={Rader, Emilee and Cotter, Kelley and Cho, Janghee},
  booktitle={Proceedings of CHI},
  pages={103},
  year={2018},
  organization={ACM}
}

@article{floridi2018ai4people,
  title={AI4People -- An ethical framework for a good AI society: opportunities, risks, principles, and recommendations},
  author={Floridi, Luciano and Cowls, Josh and Beltrametti, Monica and Chatila, Raja and Chazerand, Patrice and Dignum, Virginia and Luetge, Christoph and Madelin, Robert and Pagallo, Ugo and Rossi, Francesca and others},
  journal={Minds and Machines},
  volume={28},
  number={4},
  pages={689--707},
  year={2018},
  publisher={Springer}
}

@inproceedings{stoyanovich2016data,
  title={Data, responsibly: Fairness, neutrality and transparency in data analysis},
  author={Stoyanovich, Julia and Abiteboul, Serge and Miklau, Gerome},
  booktitle={EDBT},
  year={2016} 
}

@article{drosou2017diversity,
  title={Diversity in big data: A review},
  author={Drosou, Marina and Jagadish, HV and Pitoura, Evaggelia and Stoyanovich, Julia},
  journal={Big data},
  volume={5},
  number={2},
  pages={73--84},
  year={2017},
  publisher={Mary Ann Liebert, Inc}
}

@book{DBLP:series/dcsa/BatiniS16,
  author    = {Carlo Batini and
               Monica Scannapieco},
  title     = {Data and Information Quality - Dimensions, Principles and Techniques},
  publisher = {Springer},
  year      = {2016}
}
\end{document}